\documentstyle[preprint,aps,epsf]{revtex}
\begin{document}
\preprint{GTP-96-02}
\draft
\title{\bf{Effect of Anharmonicity on the WKB Energy Splitting in a \\  
Double Well Potential} }  
\author{Chang Soo Park$^a$, Soo-Young Lee$^b$, Jae-Rok Kahng$^b$, 
Sahng-Kyoon Yoo$^{c}$, D.K. Park$^d$, \\
C.H. Lee$^e$, and Eui-Soon  Yim$^f$ }
\address{$^a$Department of Physics,  Dankook University,  Cheonan 330-714, 
Korea\\
$^b$Department of Physics, College of Science, Korea University, 
Seoul 136-701, Korea\\     
$^c$Department of Physics, Seonam University, Namwon, Chunbuk 590-170, Korea\\
$^d$Department of Physics, Kyungnam University, Masan 631-701, Korea\\
$^e$D $\&$ S Dept., R $\&$ D Center, Anam Industrial CO., LTD, Seoul 133-120, 
Korea\\ 
$^f$Department of Physics, Semyung University, Chechon 390-230, Korea}
\date{\today}
\maketitle
\begin{abstract}
We investigate the effect of anharmonicity on the WKB approximation 
in a double well potential. By incorporating the anharmonic perturbation 
into the WKB energy splitting formula we show that the WKB approximation 
can be greatly improved in the region over which the tunneling is appreciable.  
We also observe that the usual WKB results can be obtained 
from our formalism as a limiting case in  
which the two potential minima are far apart. 
\end{abstract}

\pacs{ }
\narrowtext
It is well known that quantum tunneling leads to a splitting of degenerate 
energy levels in a symmetrical two-well  potential. 
There are three approaches to the calculation of this energy splitting: 
the WKB approximation, the instanton method,   
and numerical calculation. From the comparison of the results from the WKB 
and instanton methods with those of numerical calculations, 
it was shown that the instanton method is better than 
the WKB approximation\cite{Ban}, because the WKB method is generally 
believed to have inherent errors associated with the connection 
formula\cite{Gil}. The modified-barrier \cite{Chil} and 
modified-well\cite{Bhat} formalisms have been proposed for 
the improvement of the WKB approximation. Recently the authors in 
Ref.\cite{Fri} have shown that a careful account of the phase changes 
in connection formula improves the accuracy of the WKB wave function.

In this letter we propose another formalism whereby the energy splitting 
within the WKB approximation becomes consistent with the instanton result. 
Unlike many of the WKB formalisms, the present work incorporates 
the anharmonicity into the WKB formalism, which gives a more realistic model, 
and hence more improved energy splitting result. 
In other words, the incorporation of anharmonicity results in a
level shift due to the perturbation in each well.

Consider a particle of mass $m$ in a one-dimensional symmetrical  
two-well potential 
\begin{equation}
V(x) =   \frac{m{\omega}^2}{8a^2}  (x-a)^2    (x+a)^2   ,  
\end{equation}   
where $\omega$ is the angular frequency in each well when  the two  wells are  
far apart, and $\pm  a$ are the positions of the two potential minima. 
For a tunneling to occur the separation between the two minima should be  
large enough so that the height of the barrier $\frac{m{\omega}^2 a^2  }{8}$  
is higher than the lowest energy level in each well. 
In the limit $a \rightarrow \infty $, the potential is divided into
two independent harmonic oscillator potentials in which the lowest energies  
are the same and given by $E_0   =   \frac{1}{2}\hbar  \omega $.  
When these two harmonic oscillator potentials approach each other, 
they become an anharmonic potential, so that the lowest energies are no 
longer $E_0$ because of the anharmonic perturbation.

To evaluate the lowest perturbation energies we expand $V(x)$ around
each minima $\pm a$. Since the potential is symmetric, we consider one of 
either positions. For the minimum at $x=a$ we have 
\begin{equation}
V(x) =  \frac{m{\omega}^2}{2} (x-a)^2    \left[  1 +  \frac{x-a}{a} +  \frac{3(x-a)^2 }
{a^2  }  \right].
\end{equation}
Following a standard perturbation theory it is straightforward to show that  
the perturbation energy to second order correction is   
\begin{equation}
E=E_0 \left[ 1 + \epsilon(\eta) \right],
\end{equation}
where $\epsilon(\eta)$ is defined as 
\begin{displaymath}
\epsilon(\eta) = \frac{{\eta}^2}{16} (25 - 189{\eta}^2),  
\end{displaymath} 
and we have introduced a dimensionless parameter  
\begin{displaymath}
\eta   =  \sqrt{\frac{\hbar}{m{\omega}a^2    }}  
\end{displaymath}
which is small for large $a$. We see that the first term in Eq.(3) 
corresponds to the lowest energy of the unperturbed harmonic potential  
and $\epsilon(\eta)$ is the correction term which was ignored 
in the previous studies. In the following, we demonstrate that 
this correction term plays an important role in the improvement of
the WKB approximation.

Using Eq.(3) we write the WKB level splitting formula as\cite{Lan}   
\begin{equation}    
{\Delta}E_{WKB} = \frac{2\hbar}{T} e^{-S},  
\end{equation} 
where 
\begin{eqnarray}
S   = \frac{1}{\hbar}  \int_{-\alpha}^{\alpha}   \sqrt{2m(V(x) -   E)}    dx,  
\nonumber\\  
T  =  \int_{\alpha}^{\gamma}  \frac{\sqrt{2m}}{\sqrt{E -    
V(x)}} dx,   
\end{eqnarray}
and $\pm\alpha$, $\pm\gamma$ are the four classical turning points (Fig. 1) 
corresponding to the perturbed energy $E$. $\alpha$ and $\gamma$ can be  
expressed in terms of $\epsilon(\eta)$, repectively, as  
\begin{equation}
\alpha = a \sqrt{1 - 2\eta\sqrt{1+\epsilon(\eta)}} , \hspace {10mm}  
\gamma = a \sqrt{1 + 2\eta\sqrt{1+\epsilon(\eta)}}.
\end{equation}
Since we are interested in the region with large values of $a$, 
the elliptic integrals in Eq.(5) can be performed asymptotically for 
small $\eta$. Keeping only the dominant terms in $\eta$, 
we obtain the WKB energy splitting  
\begin{equation}
{\Delta}E_{WKB} \approx \left[ {\hbar}\omega \frac{4\sqrt{e}}{\pi\eta} 
e^{ -\frac{2}{3{\eta}^2}} \right] \delta (\eta), 
\end{equation}
where $\delta(\eta)$ is defined as
\begin{equation}
\delta(\eta) = \frac{1}{\sqrt{1 + \epsilon(\eta)}} \exp \left[ \frac{\epsilon(\eta)}{2}   
- \epsilon(\eta) \ln \left( \frac{\eta \sqrt{1 + \epsilon(\eta)}}{4}\right)\right].
\end{equation}
Comparing this with the instanton result\cite{Klei}
\begin{displaymath}
{\Delta}E_{in} = \frac{4\hbar \omega}{\sqrt{\pi}\eta} e^{-\frac{2}{3{\eta}^2}},
\end{displaymath}
we find that 
\begin{equation}
\frac{{\Delta}E_{WKB}}{{\Delta}E_{in}} = \sqrt{\frac{e}{\pi}}\delta(\eta).
\end{equation}
In the limit that the two potential minima are completely separated, 
which implies $\eta \rightarrow  0$, we see from Eq.(8) that 
$\delta(\eta) \rightarrow  1$. In this regime the Eq.(9) reduces to 
\begin{equation}
\frac{\Delta E_{WKB}^{(0)}}{\Delta E_{in}} = \sqrt{\frac{e}{\pi}},
\end{equation}
where $\Delta E_{WKB}^{(0)}$ is the WKB energy splitting obtained 
without considering the anharmonicity effect. 
The ratios in Eqs. (9) and (10) as a function of dimensionless parameter 
$\eta$ are shown in Fig.2. Note that, in the range of an appreciable tunneling 
probability (e.g.,  $ 0.1 \leq \eta \leq  0.15 $), the WKB result 
obtained from the present formalism is arbitrarily close to the instanton 
result.

A few comments are addressed in the following.
Eq.(10) agrees well with the results of Refs.[1,2] in which  
the difference between the WKB approximation and the instanton method is 
claimed to be attributed to the errors introduced by the WKB connection 
formula. We note here that they\cite{Ban,Gil} obtained the energy splitting 
in the limit $a \rightarrow \infty$, which corresponds to zero tunneling
probability and is not allowed in the calculation of the energy splitting 
due to tunneling. In order for the tunneling probability not to vanish, thus, 
the two potential wells should not be far apart. 
In this case, the coupled potential wells
are simulated as an anharmonic potential more realistically than two
idealized harmonic potentials which was assumed in their calculations.
Our result shown in Eq.(7), which includes the effect of anharmonicity, 
is based on the formalism that includes the region where the tunneling 
probability is appreciable. As we can see in Table I, within the range of 
the occurence of a considerable tunneling, significant improvement 
in the usual WKB approximation can be achieved by the incorporation of the 
anharmonicity.

In summary, we suggest a more realistic formalism with anharmonicity included 
than the previous ones only with ideal harmonicity. While the results from 
Refs.[1,2] are valid only in the limiting case of $ a \rightarrow \infty$, our
approach is applicable to the broader range of the separation between two wells, 
over which the tunneling amplitude is conspicuous. Moreover, in this region, 
our formalism greatly improves the usual WKB methods, and the WKB energy 
splitting obtained from this formalism is shown to be in good agreement with 
that from the instanton method. Taking the limit $ a \rightarrow  \infty$ 
of our result, we found that our expression reduces to the previous one 
as shown in Eq.(10)\cite{Gil}. Whereas it was only conjectured 
in Ref.\cite{Gil} that the modification factor $\sqrt{\frac{e}{\pi}}$  
may come from the connection formula of the WKB approximation, we, here, 
obtained that factor through the analytical method with 
anharmonicity incorporated. 

Some of the authors would like to approciate the partial support by
Nondirected Research Fund, Korea Research Foundation, 
'93 and '95(E.S.Y and S.K.Y), and by Korea Science and Engeneering
Foundation (961-0201-005-1)

\begin{figure}
\caption{One dimensional anharmonic double well potential. $E$ is the 
perturbation energy to second order and $\pm \alpha$, $\pm \gamma$ are 
the classical turning points corresponding to $E$.}
\label{fig1}
\end{figure}

\begin{figure}
\caption{The ratios in (9) and (10) as a function of $\eta$ are plotted. 
The region of large $\eta$ has been excluded in this plot because the two 
equations (9) and (10) are asymptotic expressions for small $\eta$. In the  
limit $\eta \rightarrow 0$ (that is, $a \rightarrow \infty$) the two plots 
exactly agree with each other. }
\label{fig2}
\end{figure}

\begin{table}
\centering
\caption{A comparison between $\Delta E_{WKB}$ and $\Delta E_{in}$ within the 
range of $0.1 \leq \eta \leq 0.15 $.}
\vspace{1cm}
\begin{tabular}{dd}
{ $\eta$ } & { $\frac{\Delta E_{WKB}}{\Delta E_{in}}$ }\\
\hline
0.1   & 0.98104 \\  
0.121 & 0.99870 \\          
0.122513 & 1.00000\\
0.123 & 1.00042 \\ 
0.125 & 1.00214 \\ 
0.127 & 1.00386 \\
0.13  & 1.00644 \\  
0.15  & 1.02349 \\
\end{tabular}
\end{table}
\newpage
\epsfysize=20cm \epsfbox{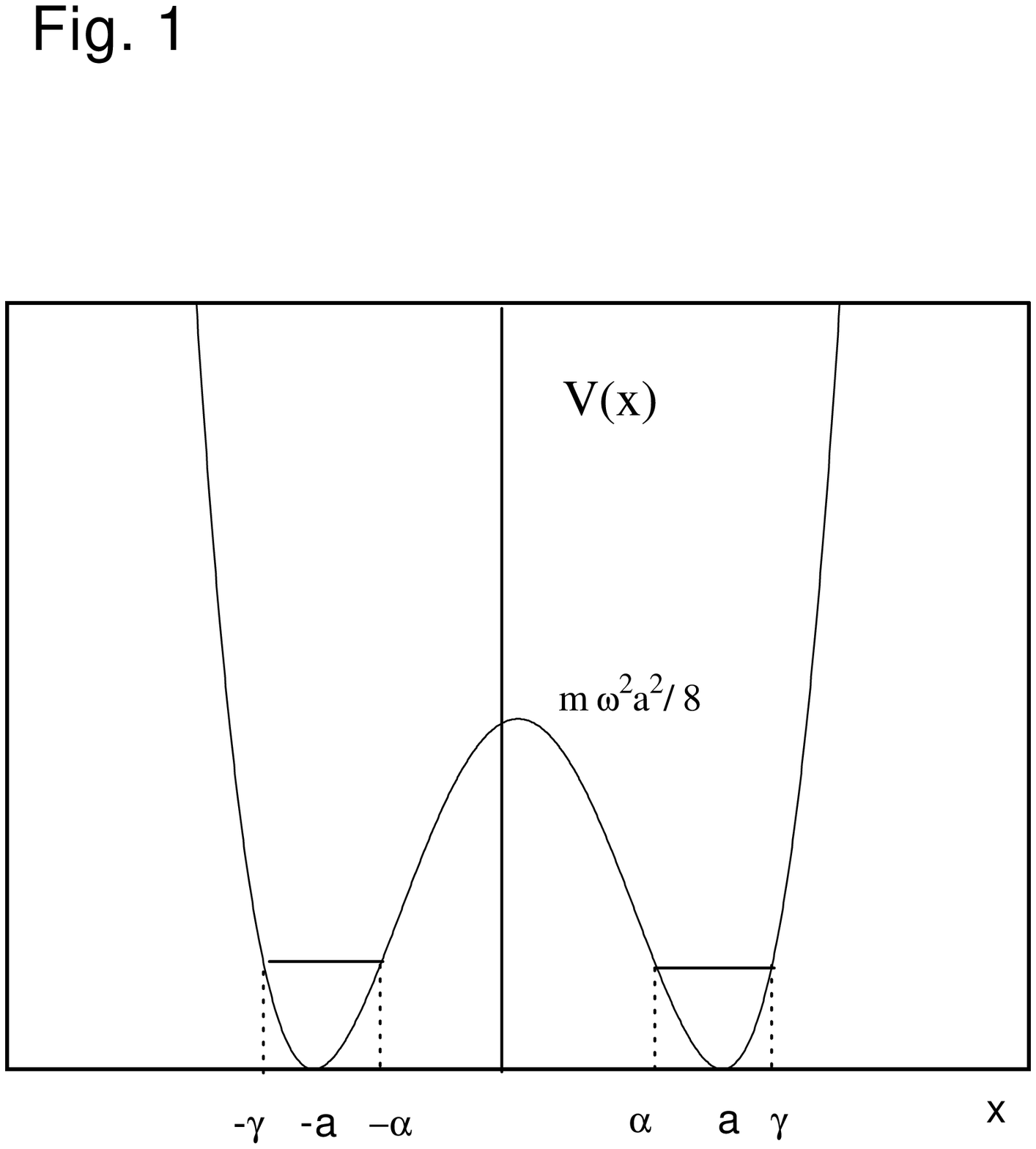}
\newpage
\epsfysize=20cm \epsfbox{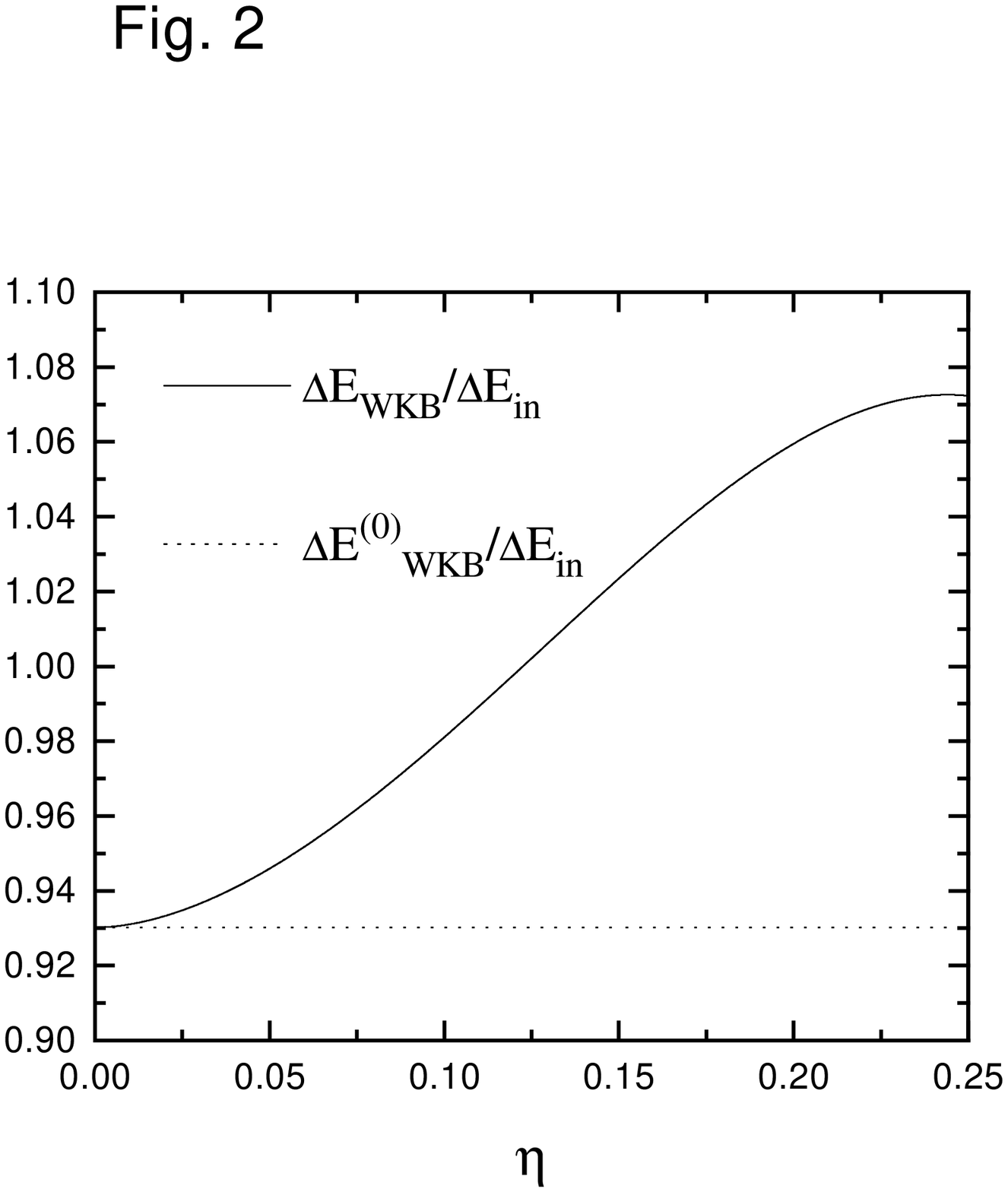}        

\begin{references}
\bibitem{Ban} K. Banerjee and P. Bhatnagar, Phys. Rev. D {\bf{18}}, 
              4767 (1978).
\bibitem{Gil} Eldad Gildener and Adrian Patrascioiu, Phys.  Rev. D 
              {\bf{16}}, 423 (1977).
\bibitem{Chil} M. S. Child, J. Mol. Spectrosc. {\bf{53}}, 280 (1974); 
               S. K. Bhattacharya and A. R. P. Rau, Phys. Rev. A {\bf{26}}, 
               2315 (1982)
\bibitem{Bhat} S. K. Bhattacharya, Phys. Rev. A {\bf{31}}, 1991 (1985).
\bibitem{Fri} H. Friedrich and J. Trost, Phys. Rev. Lett. {\bf{76}}, 4869 (1996).
\bibitem{Lan} L. D. Landau and E. M. Liftshitz, {\it{Quantum Mechanics}}, 3rd ed. 
              (Pergamon, New York, 1977), Chap. VII.
\bibitem{Klei} Hagen Kleinert, {\it{PATH INTEGRALS in Quantum Mechanics,  Statistics,
               and Polymer Physics}} (World Scientific, Singapore, 1990), Chap. 17.

\end{references}
\end{document}